\documentclass[amsmath,amssymb,prb,twocolumn,superscriptaddress]{revtex4}

\usepackage[utf8]{inputenc}
\usepackage{amsmath}
\usepackage{graphicx}
\usepackage{hyperref}
\usepackage{enumerate}

\newcommand{\dipc}{Donostia International Physics Center (DIPC), E-20018 San Sebasti\'an, Spain}

\begin{document}
\title{Olympicene radicals as building blocks of two-dimensional anisotropic networks}

\author{Ricardo Ortiz}
\affiliation{\dipc}

\date{\today}

\begin{abstract}
I propose monoradical nanographenes without $C_3$ symmetry as building blocks to design two-dimensional (2D) carbon crystals. As representative examples I study the honeycomb and Kagome lattices, showing that by replacing the sites with olympicene radicals the band dispersion near the Fermi energy corresponds, respectively, to that of Kekul\'e/anti-Kekul\'e graphene and breathing Kagome tight-binding models. As a consequence, finite islands of these new crystals present corner states close to the Fermi energy, just like the parent models. In the case of  Kekul\'e/anti-Kekul\'e graphene, such states are topologically protected, standing as examples of second-order topological insulators with a non-zero $\mathbb{Z}_2$- or $\mathbb{Z}_6$-Berry phase. Differently, those of the breathing Kagome lattice are of trivial nature, but the ground state has been predicted to be a spin liquid in the antiferromagnetic Heisenberg model. Hence, 2D systems made of low-symmetric nanographenes may be convenient platforms to explore exotic physics in carbon materials.
\end{abstract}
\maketitle

The design, synthesis and characterization of low-dimensional forms of carbon have been a priority for researchers in the past decades\cite{C60synth,nanotubes1991,graphene2004}. Considering the numerous possibilities, open-shell nanographenes were among the most challenging because of the high-reactivity of unpaired electrons, and just recent on-surface synthesis experiments has offered the necessary control for obtaining islands\cite{olympicenesynth,wang2016,pavlivcek2017synthesis,rhombenes,li2019,mishra19b,mishra2020,nonazethrene,heptauthrenesynth,NachoNanostar2021,shiyongcolab} and ribbons\cite{Ruffieux16,armchairfasel2017,armchairnacho2017,toporibbons,sun2020coupled,crommiescience} with pristine edges. However, whilst the characterization of either molecules or one-dimensional systems\cite{mishra2021observation} has been succesful, only 2D networks of open-shell nanographenes with heteroatoms\cite{C1CC12490K,2Dheterotriang} or low crystallinity\cite{2Dtriangexp} exist beyond theory\cite{topojunctionsKoshino,Ortiz_2023,zhou2020realization,sethi2021flat,liu2021designing,PRMOrtiz2022}. In this Letter, I will present three new 2D carbon crystals that are well described by model Hamiltonians where exotic physics is expected\cite{mizoguchi2019higher,lee2020fractional,hou2007electron, liu2019helical,schaffer2017quantum,repellin2017stability,iqbal2018persistence}, showing the interest inherent in nanographenes as building blocks of periodic systems.

Nanographenes with only rings formed by an even number of sites are classified as bipartite, where $N_{A,B}$ count the sites that belong to the $A$ and $B$ sublattices. When there is an imbalance between $N_A$ and $N_B$, it is impossible to accomodate all the double bonds without leaving unpaired electrons that appear as zero energy states ($N_Z$) in the non-interacting spectrum\cite{sutherland1986localization,ortiz19} ($N_Z\geq|N_A-N_B| $). Such non-bonding electrons are strongly localized at the majority sublattice, originating local moments in the presence of interactions\cite{JFR07,Yazyev_2010}, and permitting the realization of magnetism\cite{wang2008graphene} with no need of $d,f$-electrons. 


\begin{figure}[h!]
 \centering
    \includegraphics[width=0.45\textwidth]{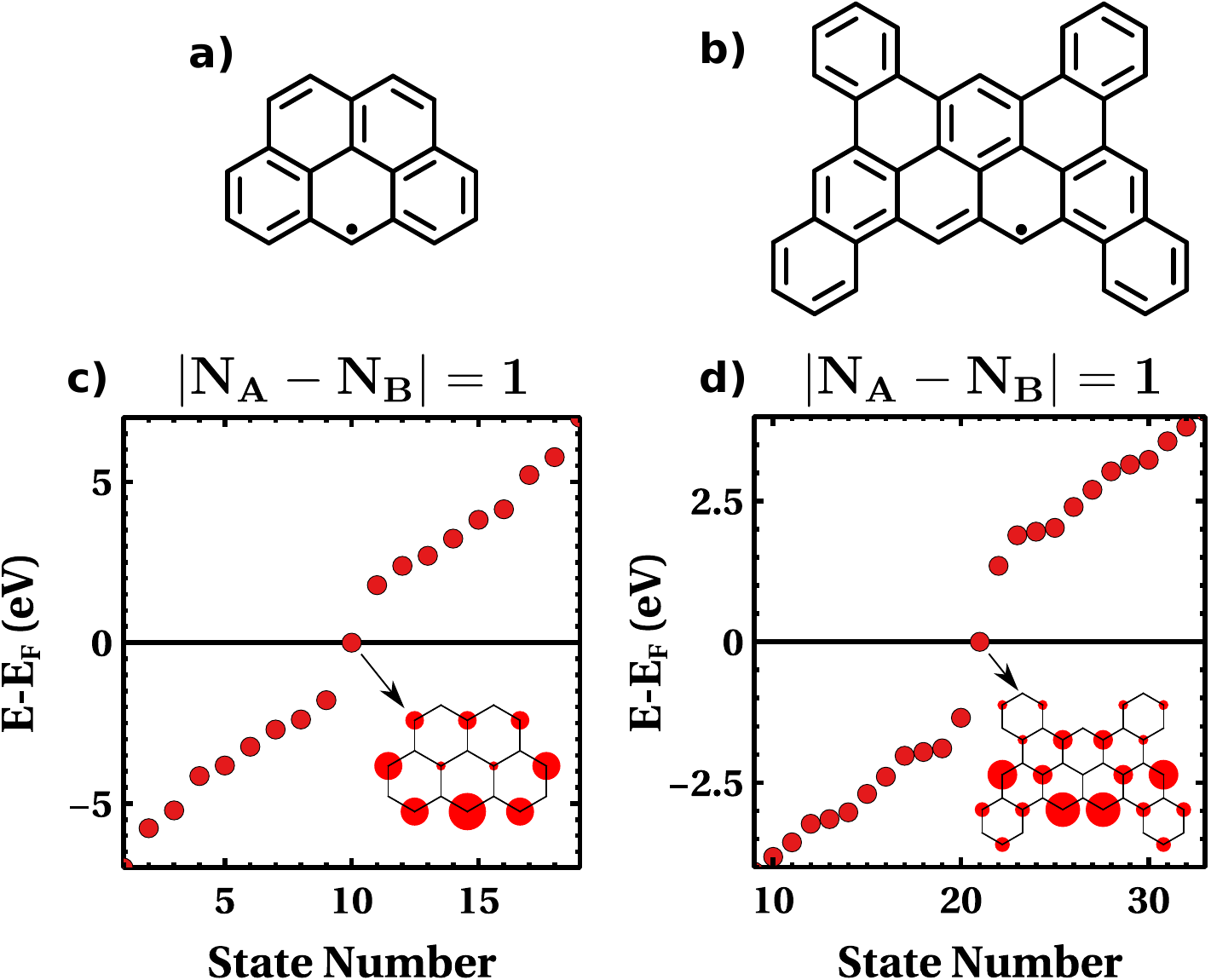}
\caption{a), b) Chemical structure of Ol and EOl molecules. c), d) First-neighbours TB eigenvalue spectra near the Fermi energy of Ol and EOl, respectively, with $t=-2.7$eV. Insets are the probability distribution of the zero energy state ($\phi_z$).}
\label{fig1}
\end{figure}

States with $|E|>0$ are usually separated by a big gap from $E=0$ states ($\delta/2>1eV$). Hence, when radical graphene fragments are covalently linked, we may often consider negligible the hybridization of the states out of the gap with those pinned at the Fermi energy\cite{Ortiz_2023} ($E_F=0$). As a consequence, the latter behave effectively as single orbitals when arranged in a structure of higher dimensionality. We can take as an example a 2D honeycomb crystal made of triangular nanographenes\cite{zhou2020realization,sethi2021flat,liu2021designing,Ortiz_2023}, known as [$n$]triangulenes, where $3n$ is the number of zigzag carbon atoms and $|N_A-N_B|=n-1$. Thus, in a spin unpolarized approximation, a 2D honeycomb network of [$2$]triangulenes (phenalenyls\cite{smalltriang,superexchangejacob}) is artificial graphene with gapless Dirac cones\cite{zhou2020realization,Ortiz_2023}, but with a smaller bandwidth because the effective hopping between two zero energy states is a fraction of the first-neighbours C-C hopping ($|\tilde{t}|<|t|$).

\begin{figure}
 \centering
    \includegraphics[width=0.48\textwidth]{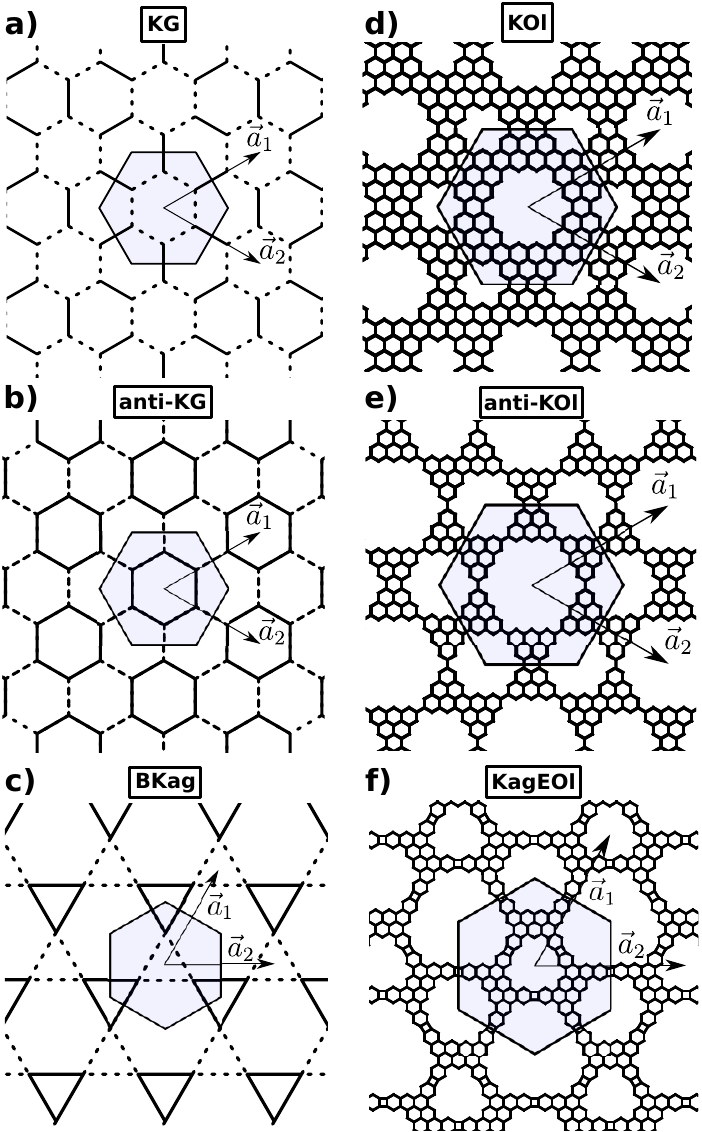}
\caption{Crystalline structures of a)-c) the model anisotropic Hamiltonians and d)-f) the three nanographene crystals under study.
The blueish hexagons are the unit cells of each crystal, and $\vec{a}_{1,2}$ are the lattice vectors. The dashed and solid lines in panels a)-c) account for weaker and stronger bonds, respectively.}
\label{fig2}
\end{figure}

In a 2D phenalenyl crystal the isotropic coupling between the phenalenyls is due to the $C_3$ rotational invariance of this molecule. In the following I will show that by selecting a building block that lacks this symmetry (Fig. \ref{fig1}) we may induce anisotropy in the effective hoppings of the zero energy states, forming new materials that can be described by 2D models with an anisotropic factor ($|t_{\mathrm{intra}}/t_{\mathrm{inter}}| \neq 1$). Among the several possibilities, the three selected examples are: Kekul\'e graphene (KG, Fig. \ref{fig2}a), anti-Kekul\'e graphene (anti-KG, Fig. \ref{fig2}b) and the breathing Kagome lattice (BKag, Fig. \ref{fig2}c); where the first two are expected to be second-order topological insulators\cite{mizoguchi2019higher,lee2020fractional}, and for the latter the ground state is predicted to be a spin liquid in the strong coupling limit\cite{schaffer2017quantum,repellin2017stability,iqbal2018persistence}. 

In this Letter three things are calculated to prove the correspondence between the systems from left and right panels of Fig. \ref{fig2}.
First, the spin unpolarized bands of the carbon crystals and models\cite{mizoguchi2019higher,bolens2019topological} showed a fair similarity, anticipating comparable low-energy physics. Second, the crystals from Fig. \ref{fig2}d (e) presented the same second Stiefel-Whitney number $w_2$ and non-zero $\mathbb{Z}_Q$-Berry phase ($\gamma_Q$) as Kekul\'e (anti-Kekul\'e) graphene\cite{mizoguchi2019higher,lee2020fractional}, so they are topologically similar: $w_2=1$ ($w_2=0$) and $\gamma_2=\pi$ ($\gamma_6=\pi$). Third, finite islands host corner states, with or without topological protection, as it is the case of the models\cite{mizoguchi2019higher,lee2020fractional,ezawa2018higher,jung2021exact,herrera2022corner,van2020topological}.


The results of this manuscript are obtained with different levels of theory. First, a tight-binding (TB) model with just one $p_z$ orbital per atom and first-neighbours hopping ($t=-2.7$eV). Particularly, for the crystal from Fig. \ref{fig2}e an additional third-neighbours hopping ($t_3=0.12t$) is included to hybridize the, otherwise disconnected, zero energy states. Then, for the fully interacting picture, Density Functional Theory (DFT) calculations are
done with the Quantum-Espresso package\cite{giannozzi2009quantum,giannozzi2017advanced,giannozzi2020} and the Perdew-Burke-Ernzerhof (PBE) density functional\cite{perdew1996generalized} for C and H atoms. Finally, for computing the local moments I used a collinear mean-field approximation of the Hubbard model (see supp. mat.\cite{SM} for  details).        


The olympicene radical\cite{olympicenesynth} (Ol, Fig. \ref{fig1}a) is a small nanographene with $|N_A-N_B|=1$, and therefore one zero energy state ($\phi_z$) in the TB eigenvalue spectrum (Fig. \ref{fig1}c). Different from phenalenyl's ingap state, $\phi_z$ does not diagonalize the $C_3$ rotation operator.
Because of this, there are just two equivalent positions (instead of three) that lead to the same hopping matrix element between $\phi_z$ orbitals of two Ol molecules:

\begin{equation}
\tilde{t}\equiv\langle\phi_{z,1}|{\cal H}_t|\phi_{z,2}\rangle,
\end{equation}
 
\noindent where ${\cal H}_t$ is the TB Hamiltonian connecting nanographenes $1$ and $2$.

A unit cell made of Ols then results in a 2D periodic system with different effective inter- and intracell hoppings ($|\tilde{t}_\mathrm{inter}|\neq|\tilde{t}_\mathrm{intra}|$).
The case of honeycomb lattices made of Ols will be analyzed, where I study two possible relative orientations between the Ol in a unit cell of six molecules: a crystal where the Ols are linked by two (three) covalent bonds with the adjacent Ols that are in (out) the unit cell (KOl, Fig. \ref{fig2}d), and a crystal made of Clar's goblets that differ in $60^\circ$ in parallel to the plane (anti-KOl, Fig. \ref{fig2}e).      

\begin{figure}
 \centering
    \includegraphics[width=0.48\textwidth]{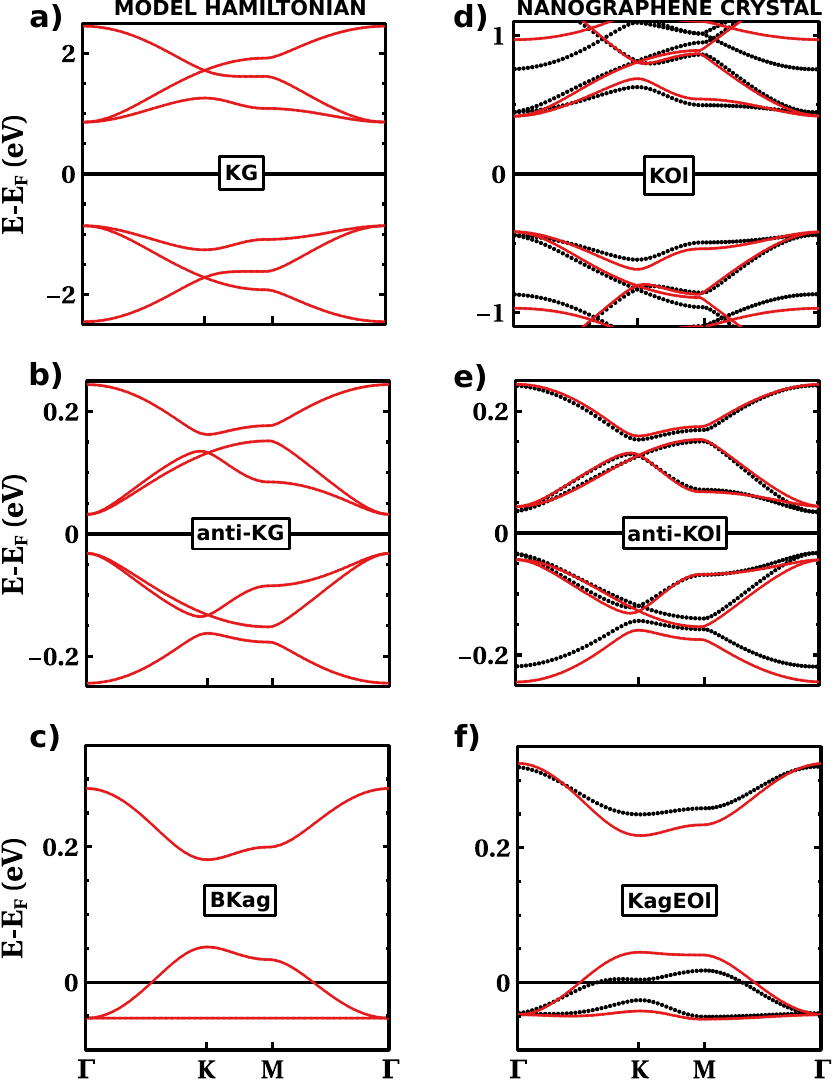}
\caption{Spin-unpolarized band structure calculated with a TB model for a) KG, b) anti-KG, c) BKag and (red solid line) d) KOl, e) anti-KOl, f) KagEOl. 
Just a first-neighbour hopping $t=-2.7$eV was considered for the nanographene crystals, except for the system of panel e) where a third-neighbour hopping was included ($t_3=0.12t$). For the effective models the hoppings where chosen to match $\tilde{t}_\mathrm{inter}$ and $\tilde{t}_\mathrm{intra}$ of the nanographene systems ($t_\mathrm{inter}=-1.39$eV and $t_\mathrm{intra}=-0.53$eV for KG, $t_\mathrm{inter}=-60$meV and $t_\mathrm{intra}=-92$meV for anti-KG, $t_\mathrm{inter}=78$meV and $t_\mathrm{intra}=35$meV for BKag). In d)-f) the black dots are the spin unpolarized band structure for the same crystals calculated with DFT. }
\label{fig3}
\end{figure}    

In the first one of them (KOl), the effective intercell is stronger than the intracell hopping ($|\tilde{t}_\mathrm{inter}|>|\tilde{t}_\mathrm{intra}|$), being described by a KG model Hamiltonian\cite{hou2007electron,lee2020fractional} (Fig. \ref{fig2}a). Adding a Kekul\'e anisotropy to the graphene TB model opens up a gap in the otherwise gapless band structure (Fig. \ref{fig3}a), which then holds a good correspondence with the first TB bands of the nanographene crystal (Fig. \ref{fig3}d). It can be seen how the first-neighbours coupling between the zero energy states causes $\tilde{t}$ to be a considerable fraction of $t$, making the upper bands to cross higher-energy bands. In addition there is a quantitative mismatch with KG for $t_\mathrm{inter}(t_\mathrm{intra})=\tilde{t}_\mathrm{inter}(\tilde{t}_\mathrm{intra})$, since the contribution of $E\neq0$ states from the Ols to the direct gap might not be totally negligible. 

Next I study the case of the 2D crystal of Clar's goblets, where it happens the opposite hopping relation ($|\tilde{t}_\mathrm{inter}|<|\tilde{t}_\mathrm{intra}|$) and the effective model is anti-KG (Fig. \ref{fig2}b). As before, hopping anisotropy opens up a gap (Fig. \ref{fig3}b), but the band dispersion is different than with the Kekul\'e distorsion\cite{mizoguchi2019higher}. The TB band structure of the anti-KOl crystal is very similar to that of this model (Fig. \ref{fig3}e). In this case, the band structures are in a better quantitative agreement when the hoppings of the anti-KG model match those of $\tilde{t}_\mathrm{intra}$ and $\tilde{t}_\mathrm{inter}$ of anti-KOl, as a consequence of a smaller hybridization between the different $\phi_z$ since they are coupled just by $t_3$.       

For the third system I tried to design a 2D carbon crystal that would model the Kagome lattice with a breathing hopping anisotropy (Fig. \ref{fig2}c). A reasonable first attempt would be to link three Ol that form the unit cell of a triangular lattice. However, the obtained non-interacting band structure did not resemble to anything remotely similar to the bands of BKag\cite{bolens2019topological}, which should be a flat band followed by a gapped Dirac cone at $K$ (Fig. \ref{fig3}c). This is undoubtedly caused by the large hybridization between $\phi_z$ states, since artificially decreasing the hoppings connecting the Ol molecules reveals the, otherwise hindered, Kagome-like bands (Fig. ~S1).

In order to correctly find a 2D nanographene with the same band structure than the fermionic BKag model, we need to make the effective hoppings smaller in a more realistic way: for instance, by shrinking the wavefunction coefficients at the linking sites. This can be done by selecting a monoradical with more C atoms than Ol, like an Ol molecule but with acene units as "legs" (Fig. ~S2). However, the unit cell now is not planar because of a steric hindrance between two H atoms, which can be avoided by enlarging the Ol skeleton in the horizontal direction that leads to the monoradical from Fig. \ref{fig1}b,d (hereafter, extended-Ol or EOl), correctly keeping the planarity of the Kagome-like crystal. This 2D carbon system (KagEOl, Fig. \ref{fig2}f) finally presents a TB band structure with a flat band and a gapped Dirac cone (Fig. \ref{fig3}f), in good agreement with the BKag model.

To show the robustness of these results I further compute the interacting spin unpolarized band structures for KOl, anti-KOl and KagEOl employing DFT and the PBE exchange-correlation functional. The geometries were properly relaxed under the BFGS quasi-Newton algorithm, with no deviation from planarity, which justifies the previous single-orbital per atom approximation. In Fig. \ref{fig3}d-f (black dots) the calculated bands are clearly comparable with those from the TB model (red lines), and therefore confirm that their low-energy physics can be anticipated by the anisotropic models from Fig. \ref{fig2}a-c.          

KG and anti-KG\cite{mizoguchi2019higher,lee2020fractional} are predicted to be second-order topological insulators\cite{benalcazar2017quantized,langbehn2017reflection,schindler2018higher} (SOTI) with ($D-2$)-dimensional modes, i.e., corner states. In inversion-symmetric spinless 2D models there is a $\mathbb{Z}_2$ invariant ($w_2$) that characterizes  SOTI, known as the second Stiefel-Whitney number\cite{ahn2018band,ahn2019symmetry}:

\begin{equation}
(-1)^{w_2}=\prod^{4}_{i=1}(-1)^{\lfloor N^-_{occ}(\Gamma_i)/2\rfloor},
\end{equation}    

\noindent where $\lfloor\rfloor$ is the floor function and $N^-_{occ}$ is the number of occupied bands with odd parity in the four time reversal invariant momenta ($\Gamma_i$), which are $\Gamma$ and three $M$ points.

\begin{figure*}
 \centering
    \includegraphics[width=0.98\textwidth]{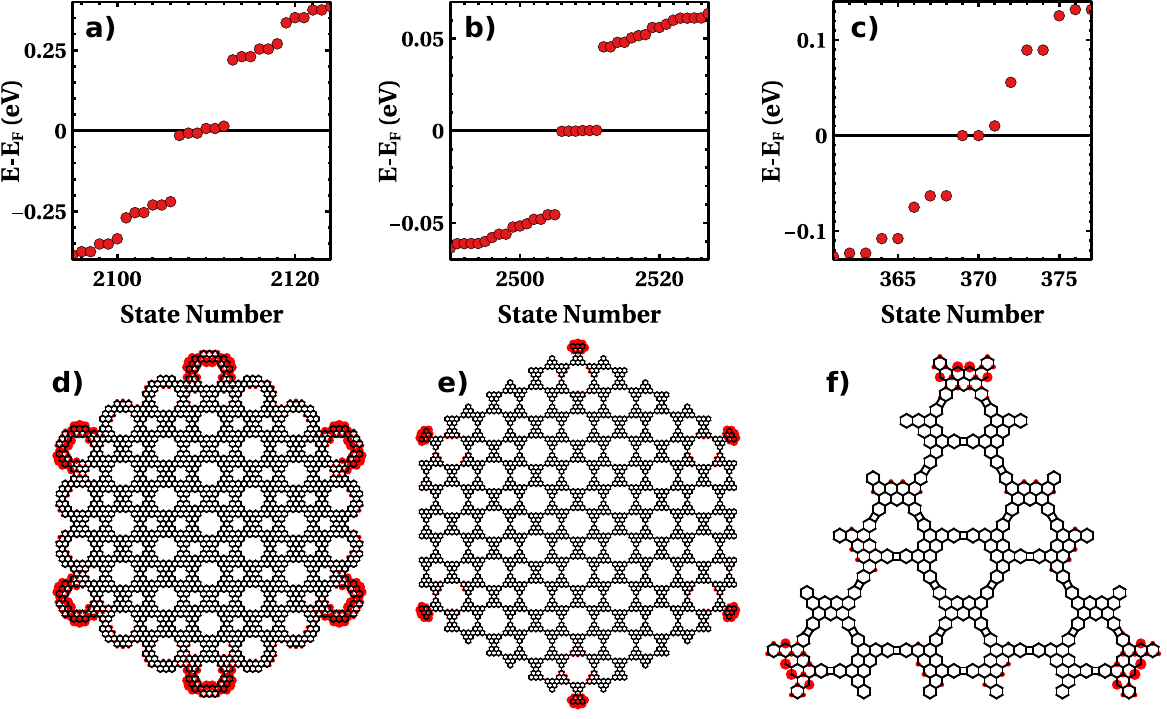}
\caption{a)-c) Single-particle spectra calculated with the TB model for finite islands of KOl, anti-KOl and KagEOl, respectively. d)-f) Propability distribution of the corner states. d) and e) correspond to the first of the six ingap states in panels a) and b). f) is the first state above $E_F$ of panel c), which is $E_F\approx-2$meV. In the three cases a first-neighbours hopping $t=-2.7$eV was considered. For panels b) and e) an additional $ t_3=0.12t$ was included. At half-filling, two electrons in total populate the corner states at $E_F$ in c).}
\label{fig4}
\end{figure*}

Materials with non-zero $w_2$ are called Stiefel-Whitney insulators (SWI)\cite{sheng2019two,pan2022two,ni2022organic}. This $\mathbb{Z}_2$ invariant can be easily calculated from the parity eigenvalues of the bands at half-filling, and KG is correctly predicted to be a SWI\cite{lee2020fractional} since a band inversion\cite{liu2017topological} at $\Gamma$ leads to $N^-_{occ}(\Gamma)=0$, $N^-_{occ}(M)=2$ and $w_2=1$. On the other hand, anti-KG has no band inversion, therefore $\Gamma$ and $M$ present the same number of occupied odd-parity bands $N^-_{occ}(\Gamma,M)=2$, that yields $w_2=0$, and in principle it should be a trivial insulator\cite{lee2020fractional}. Similarly, I compute the parities of the occupied bands of KOl and anti-KOl, where $N^-_{occ}(\Gamma)=27$ and $N^-_{occ}(M)=29$ for the former, and $N^-_{occ}(\Gamma,M)=29$ for the latter. This leads to $w_2=1$ and $w_2=0$, respectively, in good correspondence with the two anisotropic graphene models. 

According to the previous paragraph, KOl (anti-KOl) is predicted to be a Stiefel-Whitney (trivial) insulator, like KG (anti-KG) if we attend to $w_2$. However, it has also been suggested in several works\cite{mizoguchi2019higher,araki2020z,bunney2022competition} the use of the $\mathbb{Z}_Q$-Berry phase ($\gamma_Q$) as a topological invariant to characterize SOTI phases: 
\begin{equation}
\gamma_Q=\int_{L}d\bf{\Theta}\cdot \bf{A}(\bf{\Theta}),
\end{equation}

\noindent where the integral goes over a path $L$ in a $Q-1$ parameter space ${\bf{\Theta}}=(\Phi_1,\Phi_2...\Phi_{Q-1})$, $\bf{A}(\bf{\Theta})$ is the Berry connection:

\begin{equation}
{\bf{A}}({\bf{\Theta}}) = -i\langle{\bf{\Psi}}({\bf{\Theta}})|\partial_{{\bf{\Theta}}}{\bf{\Psi}}({\bf{\Theta}})\rangle,
\end{equation}

\noindent and ${\bf{\Psi}}({\bf{\Theta}})$ is the many-body ground state of a non-interacting Hamiltonian $H({\bf{\Theta}})=h_0({\bf{\Theta}})+(H-h_0) $ with a local twist described by $t_j\rightarrow t_je^{i\Phi_j}$. This Hamiltonian is defined on a finite system with periodic boundary conditions (PBC), where $h_0$ contains the bonds ($t_j$) that will have the twist in $h_0(\bf{\Theta})$. Particularly, in the anisotropic graphene models $h_0$ is the part of an individual cluster from a cluster-limit (see supp. mat.\cite{SM} or references\cite{mizoguchi2019higher,araki2020z,bunney2022competition} for more details).

Such Berry phase allows the characterization of different non-trivial topological phases with cluster-limits that differ in symmetry\cite{mizoguchi2019higher,bunney2022competition}.  
For instance, for each anisotropic graphene model we can define a cluster-limit setting the weaker bond to zero, that consists in a set of disconnected dimers or hexamers (Fig~S3). The $C_2$ or $C_6$ symmetries from these clusters permit the quantization of the Berry phase in $\mathbb{Z}_2$ or $\mathbb{Z}_6$, respectively, obtaining a calculated $\mathbb{Z}_Q$-Berry phase for KG (anti-KG) of $\gamma_2=\pi$ ($\gamma_2=0$) but $\gamma_6=0$ ($\gamma_6=\pi$). This means that both are SOTI with different limits that cannot be adiabatically transformed one into the other without closing the gap\cite{mizoguchi2019higher,bunney2022competition}. 

Analogously, for KOl and anti-KOl the local twist needs to be considered for every hopping (either $t$ or $t_3$) that connects $Q$ Ol molecules (Fig~S4). The numerical Berry phases, calculated for KOl (anti-KOl), are $\gamma_2=\pi$ ($\gamma_2=0$) and $\gamma_6=0$ ($\gamma_6=\pi$), indicating that their ground states are adiabatically connected to cluster states formed by a collection of disconnected Ol dimers (hexamers, Fig~S3). Consequently, like the anisotropic graphene models, these carbon crystals should be expected to be SOTI, and hexagonal islands will present six single-particle ingap states that are strongly localized at the corners (Fig. \ref{fig4}a,b,d,e).



It is worth to mention that the presence alone of corner states does not ensure a higher-order non-trivial topology\cite{jung2021exact}. It is needed a bulk-boundary correspondence between the bulk topology and the existence of such states. If that is the case, then there is a topological protection as long as certain symmetries are respected. We can make the appropiate tests on KG, anti-KG, KOl and anti-KOl, showing the robustness of their corresponding corner states (Figs~S5, S6, S7).
The BKag model, on the other hand, also presents corner states but fragile against perturbations\cite{herrera2022corner,van2020topological}. For some years it was thought that this model was also a SOTI, but its corner states are actually from trivial origin\cite{herrera2022corner,van2020topological}, hence the $\mathbb{Z}_3$-Berry phase is not a good topological invariant for this system\cite{jung2021exact}. In any case, I show in Fig. \ref{fig4}c,f that a finite island of KagEOl has also corner states, but different to KOl and anti-KOl, and similar to BKag, not robust against perturbations (Fig~S8).

I next briefly discuss the spin-polarized picture, which is relevant for realistic radical nanographenes prone to develop local moments\cite{Yazyev_2010,superexchangejacob,de2022carbon} depending on the $\tilde{U}/|\tilde{t}|$ ratio: 

\begin{equation}
\tilde{U}/|\tilde{t}| = U\sum_i |\phi_z(i)|^4/|\tilde{t}|,
\end{equation} 

\noindent where the sum runs over sites $i$ and $\sum_i |\phi_z(i)|^4$ is the inverse participation ratio.


\begin{table}
 \begin{center}
    \begin{tabular}{ r | r | r | r }
	\hline
	\hline
	2D system & $\tilde{U}/U$ & $\tilde{t}_\mathrm{inter}/t $ &  $\tilde{t}_\mathrm{intra}/t$ \\ \hline
	KOl &  0.14  & 0.515  & 0.196  \\
	anti-KOl & 0.14  & 0.022 & 0.034  \\
	KagEOl  & 0.09  & -0.029 & -0.013  \\
	\hline\hline
	\end{tabular}
\end{center}
\vspace{-3mm}
\caption{Effective on-site Coulomb repulsion and hopping matrix elements in the inter- and intra-cell directions for the Ol and EOl radicals in the KOl, anti-KOl and KagEOl crystals.}
\label{table1}
\end{table} 

For the KOl unit cell, zero energy states from two Ol molecules have a high hybridization both in the inter- and intra-cell directions (Table \ref{table1}), that leads to $\tilde{U}/|\tilde{t}|_\mathrm{inter}= 0.41$ and $\tilde{U}/|\tilde{t}|_\mathrm{intra}= 1.07$, with an on-site Coulomb repulsion $U=1.5|t|$. Two hybridized nanographenes preserve the open-shell character if $\tilde{U}\gg|\tilde{t}|$, breaking down the picture when $\tilde{U}$ and $|\tilde{t}|$ are comparable\cite{superexchangejacob}, so KOl is probably non-magnetic. 
Anti-KOl, on the other hand, has an intermolecular hybridization that is one order of magnitude lower because the $\phi_z$ states are just coupled by $t_3$ (Table \ref{table1}). In that case, $\tilde{U}/|\tilde{t}|_\mathrm{inter}= 9.54$ and $\tilde{U}/|\tilde{t}|_\mathrm{intra}= 6.18$, for $U=1.5|t|$, which suggests that all the Ol units have local moments. This is supported by calculations of the spin density per site done with a mean-field Hubbard model, where an hexagonal island of KOl just presents localized moments at the corners, while a finite island of anti-KOl has magnetization at every single Ol molecule for the same value of $U$ (Fig~S9).

Hatsugai and coworkers\cite{kudo2019higher} suggested the emergence of a novel phase in interacting SOTI known as the higher-order topological Mott insulator (HOTMI), characterized by the spin counterpart of $\gamma_Q$. In this topological state correlation annihilates the gapless charge excitations of a SOTI, hosting instead gapless spin excitations at the corners. Consequently, if interacting KG and anti-KG are HOTMIs (and hence also KOl and anti-KOl), free spins should appear localized at the corners\cite{kudo2019higher}.     
The particular case of the BKag lattice is also very interesting, since the presence of non-alternant rings promotes magnetic frustration, which stabilizes the emergence of a spin liquid ground state in the antiferromagnetic (AF) $S=1/2$ Heisenberg  model\cite{iqbal2018persistence,schaffer2017quantum,repellin2017stability}. The KagEOl crystal has a $\tilde{U}/|\tilde{t}|$ ratio fairly away from the $\tilde{U}\approx|\tilde{t}|$ scenario (Table \ref{table1}) because of the reduced weight of $\phi_z$ in the linking atoms ($\tilde{U}/|\tilde{t}|_\mathrm{inter}=4.66$ and $\tilde{U}/|\tilde{t}|_\mathrm{intra}=10.38$, for $U=1.5|t|$). As it happens with anti-KOl, every EOl unit has local magnetization in a mean-field Hubbard calculation. Thus, I suggest that it might be well described by a spin model. In any case, whether KOl and anti-KOl are HOTMIs, or KagEOl has actually a spin liquid ground state, would be the aim of future work. 

In conclusion, monoradical nanographenes with no $C_3$ symmetry serve as building blocks for 2D systems described by anisotropic models. The three proposed crystals, called here KOl, anti-KOl and KagEOl, held comparable spin-unpolarized band structures than KG, anti-KG and BKag, respectively, showing also corner states in finite islands. KOl and anti-KOl have the same $w_2$ and $\mathbb{Z}_{2,6}$-Berry phases than KG and anti-KG, so they are also SOTI, and form part of a growing list of carbon materials with predicted higher-order topology like graphdiyne\cite{sheng2019two,lee2020two} or graphene antidots\cite{xue2021higher}.
The non-interacting model Hamiltonians mentioned through this work have been relevant in photonic\cite{noh2018topological,el2019corner} and acoustic systems\cite{ni2019observation,yang2023characterization}, among others\cite{kempkes2019robust,fan2019elastic}, but electronic correlation is important in nanographenes that can develop local magnetic moments. According to the $\tilde{U}/|\tilde{t}|$ ratio, KOl is likely to be non-magnetic, but anti-KOl and KagEOl might be well described by spin models, thus we can expect the latter to host a spin liquid ground state with potential applications in topological quantum computing. All of these results should fuel the already increasing interest towards 2D nanographene crystals, whose general growth on surfaces may become a reality in the incoming years.


\begin{acknowledgments}
I acknowledge T. Frederiksen, G. Giedke, J. Fern\'andez-Rossier, J. C. Sancho-Garc\'ia, T. Mizoguchi and Y. Hatsugai for fruitful discussions.
This work was funded by the Spanish MCIN/AEI/ 10.13039/501100011033 (PID2020-115406GB-I00), and the European Union’s Horizon 2020 (FET-Open project SPRING Grant No.~863098).
\end{acknowledgments} 

\noindent contact e-mail: roc6493@gmail.com

\bibliographystyle{apsrev-title}
\bibliography{references.bib}

\end{document}